\documentclass[aps,prd,superscriptaddress,twocolumn]{revtex4}
\usepackage{amsmath,epsfig,mathrsfs,graphicx,amssymb,color}
\usepackage[T1]{fontenc}
\usepackage{t1enc}
\usepackage{hyperref}
\newcommand{\h}[1]{\hat{ #1}}
\newcommand{\f}[2]{\frac {#1}{#2}}
\def\e{\boldsymbol e}

\newcommand{\g}[1]{{\boldsymbol #1}}

\def\rmi{\mathrm i}
\def\rme{\mathrm e}
\def\be#1\ee{\begin{equation}#1\end{equation}}
\newcommand{\ba}{\begin{eqnarray} }
\newcommand{\ea}{\end{eqnarray} }

\begin{document}

\title{Effect of relativity and vacuum fluctuations on quantum measurement}
\author{Adam Bednorz}
\affiliation{Faculty of Physics, University of Warsaw, ul. Pasteura 5, PL02-093 Warsaw, Poland}
\email{Adam.Bednorz@fuw.edu.pl}

\author{Wolfgang Belzig}
\email{Wolfgang.Belzig@uni-konstanz.de}
\affiliation{Fachbereich Physik, Universit{\"a}t Konstanz, D-78457 Konstanz, Germany}

\date{\today}

\begin{abstract}
Vacuum fluctuations can obscure the detection signal of the measurement of the smallest quantum objects like single particles seemingly 
implying a fundamental limit to measurement accuracy. However, as we show relativistic invariance implies the disappearance of fluctuations 
for the space-like spectrum of an observable at zero temperature. This complete absence of noise can be harnessed to perform noiseless measurement 
of single particles, as we illustrate for electrons or photons.
We outline a general scheme to illustrate the noiseless measurement involving the space-like spectrum of observables based on 
the self-interference of counter-propagating paths of a single particle in a triangular Sagnac interferometer.
\end{abstract}

\maketitle

\section{Introduction}
The standard projection postulate is insufficient to define realistic quantum measurement in fundamental relativistic quantum theories \cite{peres,breuer}. 
Relativistic measurements are often related to the fictitious Unruh-deWitt detection model \cite{unruh,witt} that is hardly feasible experimentally. Current standard high-energy measurements \cite{pdg} are realized for large beam colliders and the detection is usually destructive since the particles are absorbed.
Various theoretical models of relativistic measurements are constantly being invented \cite{rsg,rove,lin,martinez,anas,brody}.
The problem is that such invasive measurements heavily disturb the system, destroy coherence or make undesired changes in the state \cite{brag,clerk}.

To avoid problems related to the projections postulate, one could introduce weak, almost noninvasive measurements \cite{aav, bfb} that in principle allow detection of particles without losing coherence. 
The measurement is always accompanied by some noise.
There are three main sources of this noise: internal detector's noise, thermal and zero-temperature vacuum fluctuations.
The detector's noise is large for weak measurements, but
its effect can be reduced by averaging the results of many independent detectors.
In this paper we focus in particular on vacuum fluctuations in the limit of zero temperature. 
Such a vacuum noise can be much larger than the signal. For example, when detecting a single particle, the vacuum noise can make it impossible to distinguish 
between the presence and absence of the particle. Estimating the effect of vacuum fluctuations on the measurement, one has to take into account the spatio-temporal scale of the detection. In particular, the fluctuations cannot be neglected if the measurement time $\tau$ is shorter than the inverse mass $m^{-1}$ of the particle. Obviously massless particles require a separate treatment. Electrons in a vacuum have a mass largely exceeding the present measurement frequency scale (in the sense of time-energy uncertainty).  In all such cases  the actual influence of vacuum fluctuations on the detection of single particles involves many factors: detector efficiency (some particles remain undetected), dark counts (false registering of a particle when none actually arrived) or the spatial scale of the detector. 

The above questions can also  be studied in condensed matter systems with an analogous description of particles as in high energy physics. 
Here interaction effects or the band structure can lead to effective masses larger or smaller than that of free particles. 
Furthermore, the relevant velocity is the Fermi velocity, which is typically 2 or 3 orders of magnitude smaller than the velocity of light. Interestingly, the two-dimensional Dirac dynamics of relativistic electrons is realized in graphene and other 2D crystals \cite{graphene,kane}. Furthermore, Dirac semimetals feature a three-dimensional Dirac cones
\cite{dirac3da,dirac3db,dirac3dc}. A large variety of engineered systems allow constructing detectors in a regime that is inaccessible in high-energy physics \cite{neder,oliver}.

The noise can be only partially suppressed by reducing temperature because some fluctuations remain nonzero even at zero temperature. As we show below, 
the fluctuation-dissipation theorem and the positivity of $ \langle X^2\rangle $ combined with relativistic invariance imply that the spectral density of certain fluctuations must be zero
for space-like frequency-wavevector relations. As possible realization, we suggest a setup in which a particle is sent into a triangular Sagnac interferometer \cite{sagnac}. The interferometer will produce a standing wave, that can be detected by observing a physical quantity in the space-like spectrum, e.g. the components of the electric current or the energy-momentum tensor. The setup works both for fermionic as well as bosonic particles. 

The paper is organized as follows. We start by recalling the mathematical conventions in any relativistic quantum field theory. Then
we show the disappearance of vacuum fluctuations at zero temperature and space-like spectrum. Next, we outline the proposed measurement setup
based on Sagnac interferometer. Finally, we discuss possible applications and the development of the proposed scheme. Some lengthy proofs are left in Appendices.

\section{Vacuum fluctuations in the space-like spectrum}

There exists a vacuum state, which is invariant under Lorentz transformations (assuming flat metric), which is either postulated \cite{wightman,jost,coleman} or proved perturbatively \cite{ab13}. One can consider also thermal states, described by a temperature $T$, with $T=0$ being the standard vacuum, or add perturbations (e.g. single particles).
For $T\neq 0$ thermal states are defined in a preferred reference frame.
%
Formal quantum correlations of fields (operators $\hat{A}$, $\hat{B}$, $\hat{C}$,...) in the state
$\hat{\rho}$ have the form

\begin{equation}
\langle A(x)B(y)C(z)\cdots\rangle=\mathrm{Tr}\left[\hat{\rho} \hat{A}(x)\hat{B}(y)\hat{C}(z)\cdots\right]\,.\label{ccc}
\end{equation}
These correlations describe the measurable properties of a given system and correspond to correlation measurements, response functions, or internal noises. They do depend on the order of operators, as these are usually not commuting.  
In principle, every quantum process can be
written in terms of correlations (\ref{ccc}). Technically the calculation of (\ref{ccc}) involves usually closed time path (CTP) formalism \cite{keldysh,schwinger,kaba,matsubara}, see Appendix \ref{appa}, but it is important that correlations are Lorentz-invariant in the vacuum
 at zero temperature.

Let us analyze the universal relativistic properties of second-order correlations of local fields in the invariant zero-temperature vacuum or at very low temperatures (compared to relevant energy-momentum scales). For relativistic notation conventions, see Appendix \ref{appa}.
For a uniform quantity $X(x)$ (i.e., depending only on local fields, not directly on $x$)
the translation symmetry (valid also at nonzero temperature) implies
\begin{equation}
    \langle X(q)X(p)\rangle=\delta(q+p)G(p)
\end{equation}
with $X(p)=\int e^{ix\cdot p}X(x)dx$ and $G(-p)=G^\ast(p)$. Moreover, the fluctuation-dissipation theorem states that
$G^\ast(p)=e^{-\beta p^0}G(p)$ \cite{fdt} for the inverse temperature $\beta=1/T$.
In the zero-temperature limit $\beta\to +\infty$ all correlations become Lorentz-invariant (covariant if $X$ is combined
with a Lorentz vector) \cite{wightman,jost,coleman,ab13}.
For spacelike $p$ we can always find a reference frame with $p^0>0$ and then $e^{-\beta p^0}\to 0$, and accordingly $G(p)\to 0$. 
Hence, we can write  $G=G_+(p\cdot p)\delta_+(p\cdot p)$
with $\delta_+(p\cdot p)=\theta(p_0)\delta(p\cdot p)$ selecting the forward time-like cone for $p$.

In fact all two-point correlations
\begin{equation}
\langle X(p)Y(q)\rangle=\delta(p+q)G_{XY}(p)
\end{equation}
are suppressed exponentially  by the factor $\exp(-\beta (|\g{p}|-|p^0|)/2)$ in $G_{XY}$ \cite{bound1,bound2},  which we also prove perturbatively in Appendix \ref{appc} for a generic family of quantum field theories. In the limit $\beta \to +\infty$ at constant $p$ it means again that $G_{XY}(p)\to 0$ at 
$p\cdot p<0$. 
 If $G_{XX}=\langle X(p)X(-p)\rangle=0$ (and $\langle X(p)\rangle=0$) then, for a positive underlying probability, we have a general observation that
\begin{equation}
X(p)|\mathrm{vac}\rangle=0,\:p\cdot p<0, \label{noiseo}    
\end{equation} meaning that the vacuum state $|\mathrm{vac}\rangle$ is the eigenstate of $X(p)$ with the eigenvalue $0$ and so $X(p)$ does not fluctuate then at all, it is noiseless.
This universal property of relativistic field theories is the main result of the article and  fundamental for our subsequent analysis. Any observable with zero average defined within the space-like spectrum must be suppressed completely to zero in the (zero-temperature) vacuum.

\subsection{Alternative proof by symmetry and positivity}
\label{sec2a}

Interestingly, for special vector and tensor quantities, like the electric current $j^\mu(p)$ or the energy-momentum tensor
$T^{\mu\nu}(p)$, the lack of noise in the vacuum state follows purely from translation and Lorentz invariance and the assumption that second order correlations are positive definite, i.e.,\ $\langle X^2\rangle\geq 0$ for any quantity $X$ which can be spacetime-dependent or a linear combination of other quantities. 
We do not need quantum mechanics at all to show it although it is worth mentioning that all second order quantum correlations are indeed positive definite \cite{bbb}.

Let us show it for a generic real vector quantity $A^\mu(x)$.
From translation invariance we have
\begin{equation}
\langle A^\mu(p)A^\nu(q)\rangle=\delta(p+q)G^{\mu\nu}(p)\,.
\end{equation}
The Lorentz invariance implies for $p\cdot p<0$ that $G^{\mu\nu}(p)=p^\mu p^\nu \eta(p\cdot p)$  with some function $\eta(p\cdot p)\geq 0$, \cite{ab15,ab16}
(see also Appendix \ref{appd}). 
For $p=(0,...,0,p^D)$ in $D+1$ dimensional spacetime we have noiseless components $A^{\mu}$ for $\mu=0,...,D-1$. 
We note that a generic $A^\mu(p)$ can be projected onto a noiseless quantity defined by $\tilde{A}^\mu(p)=(p\cdot p)A^\mu(p)-p^\mu(p\cdot A)$. Hence, it is always possible to defined a noiseless vector observable, at least for some components. Moreover, if $A^\mu$ is a conserved quantity, i.e., $\partial_\mu A^\mu=0$, then $p\cdot A(p)=0$ so $p_\mu G^{\mu\nu}(p)=0$  and $\eta=0$.
Note that
one cannot assume conservation in the case of anomalies \cite{ad69,belljack, adbar,ber} but our general proof in Appendix \ref{appc} is then still applicable.

A bit more complicated reasoning applies to a generic symmetric tensor quantity $B^{\mu\nu}(x)=B^{\nu\mu}(x)$ Then
\begin{equation}
\langle B^{\mu\nu}(p)B^{\sigma\rho}(q)\rangle=\delta(p+q)G^{\mu\nu\sigma\rho}(p)
\end{equation}
Lorentz invariance and positivity implies that for \newline $p\cdot p<0$ (see Appendix \ref{appd}),
\begin{eqnarray}
&&G^{\mu\nu\sigma\rho}=(g^{\mu\nu}-bp^\mu p^\nu)(g^{\sigma\rho}-bp^\sigma p^\rho)w\nonumber\\
&&+p^\mu p^\nu p^\sigma p^\rho a\label{gbb}
\end{eqnarray}
with $w,a\geq 0$ and $b$ being real functions of $p\cdot p$. If additionally conservation holds, i.e., $\partial_\mu B^{\mu\nu}=0$ (e.g. in the case of energy-momentum), then $p_\mu G^{\mu\nu\sigma\rho}=0$ giving further restriction
\begin{equation}
G^{\mu\nu\sigma\rho}=(g^{\mu\nu}-p^\mu p^\nu/p\cdot p)(g^{\sigma\rho}-p^\sigma p^\rho/p\cdot p)w
\end{equation}
with $w\geq 0$. For example, taking $p=(0,0,0,p^3)$  the observables 
$B^{12}$, $B^{01}$, $B^{02}$, $B^{11}-B^{22}$, and $B^{00}+B^{11}$ are noiseless (equal $0$). Again, a generic $B^{\mu\nu}(p)$ can be projected onto a noiseless one by defining
\begin{eqnarray}
&&\tilde{B}^{\mu\nu}(p)=(p\cdot p)^2B^{\mu\nu}(p)\nonumber\\
&&-(p\cdot p)(g^{\mu\nu}(p\cdot p)-p^\mu p^\nu)B/3\\
&&
+(g^{\mu\nu}(p\cdot p)-4p^\mu p^\nu )p\cdot B\cdot p/3\nonumber
\end{eqnarray}
with $B=B^\mu_\mu$.
In the case of a conserved quantity $p\cdot B\cdot p\equiv 0$ identically so, dividing by $p\cdot p$, we have just
\begin{equation}
\tilde{B}^{\mu\nu}(p)=(p\cdot p)B^{\mu\nu}(p)-(g^{\mu\nu}(p\cdot p)-p^\mu p^\nu)B/3\,.
\end{equation}

For an antisymmetric second rank tensor $B^{\mu\nu}=-B^{\nu\mu}$ (e.g. the electromagnetic field $F^{\mu\nu}=\partial^\mu A^\nu-\partial^\nu A^\mu$) the generic invariant correlation gives
simply $G=0$ in this case (see Appendix \ref{appd}). Such tensor analysis can be certainly generalized to
more complicated tensors.  Therefore a family of generic vectors and tensors remains noiseless in the spacelike spectrum in the vacuum, just assuming invariance and positivity. 

\subsection{Approximate localization}

In practice, all noiseless observables will be localized in real spacetime, resulting in a convolution in momentum space
\begin{equation}
A^\mu(\bar{p})\to \int dp A^\mu(p)N(p-\bar{p})
\end{equation}
with some distribution $N$ (e.g. Gaussian). Then the inverse variance of $N(p)$ gives the spatiotemporal spread of $A^\mu(x)$. We can still project onto a noiseless subspace, because the multiplication by a polynomial of $p$ corresponds to derivatives in spacetime.
It is impossible to have exactly $N(p)=0$ for all timelike $p$, but it is more realistic to assume it decays exponentially for a sufficiently smooth localization function $N(x)$.
Therefore, the contribution to the noise due to the finite spacetime volume of the measurement is negligible in practice.


\subsection{Realistic interpretation}

Quantum correlations involving only two observables can be explained a classical probability (positive definite) \cite{bbb}. However, some correlations involving  three or more observables, calculated using (\ref{ccc}),  can violate the classical inequalities like Cauchy-Bunyakovsky-Schwarz inequality
$\langle XY\rangle^2\leq \langle X^2\rangle\langle Y^2\rangle$
for at least one composite observable e.g. $X=X_aX_b$. Therefore, general correlations defined by (\ref{ccc}) make a realistic interpretation by a statistical distribution with a positive probability impossible. As further consequences, higher order correlations such as $\langle XYZ\rangle$ can remain nonzero even if e.g. $X$ is noiseless in the sense of (\ref{noiseo}). For a detailed discussion of these problems we refer the readers to \cite{bfb,ab16,energy}, 
and examples in quantum field theory in Appendices \ref{appe} and \ref{appf}. Not all correlations exhibit these problems and one can restrict the set of observables to those explained by a positive probability. Another potential solution could be to define correlations in a different way from (\ref{ccc}). Even if such a definition is possible, it may be very hard within the natural class of quantum interpretations (based on measurement models within the usual quantum field theory algebra), so we leave this discussion for future considerations.

\section{Application to measurements of single particles}

In this section, we investigate the advantage  of noiseless observables in single particle detection over the standard projection. We shall see that the vacuum fluctuations, contributing to the noise of low-mass particle detectors can be eliminated by using noiseless observables and a homodyne detection using a Sagnac interferometer.

\subsection{Vacuum fluctuation in a simple measurement}

Suppose we want to measure a single particle by an observable
\begin{equation}
\bar{S}=\int_{V,\tau}S(x)dx.\label{sss0}
\end{equation}
where $S=j^\mu$, or $T^{00}$, i.e., measuring current or energy density, over the spatial volume $V$ and time $\tau$.
In the vacuum $\langle \bar{S}^2\rangle$ is determined purely by vacuum fluctuations.
For free theories, if the particle has a mass $m$ (either boson or fermion) then the minimal frequency of the vacuum fluctuations is $2m$
(similarly to the Zitterbewegung).
Therefore  $\langle \bar{S}^2\rangle=0$ for $2m\tau<1$ and the vacuum fluctuations do not mask realistic detection signals which have a frequency scale $1/\tau \ll m$.

However, the situation is different in the massless case, including low-energy systems, e.g. descibed by an effective mass in the band structure in condensed matter. Here, the contribution of the vacuum fluctuations is nonzero and can be estimated.
For e.g. current density, $S=j^1$, we can use the massless relativistic results for current-current correlations in $D$ dimensions, i.e., spacetime $(x^0,x^1,...,x^D)$
\ba
 &&\langle j^\mu(p)j^\nu(q)\rangle\sim \delta(p+q)(p^\mu p^\nu-g^{\mu\nu}p\cdot p)\nonumber\\
&&\times\left\{\begin{array}{ll}
 \delta(p\cdot p)&\mbox{ for }D=1\\
 \theta(p\cdot p)/\sqrt{p\cdot p}&\mbox{ for }D=2\,.\\
  \theta(p\cdot p)&\mbox{ for }D=3
 \end{array}
 \right.
\ea
Taking a detector of volume $V$ and measurement time $\tau\gg V^{1/D}$ the fluctuations of  $\bar{S}$ for $S=j^1$ read $\langle \bar{S}^2\rangle\sim V^2\tau^{2-2D}$. Analogously, for the energy-momentum and electromagnetic field, with $S=T^{00}$, we have $\langle \bar{S}^2\rangle\sim V^2\tau^{-2D}$, so the vacuum fluctuations of $\bar{S}$ are still finite. Note that these fluctuations are algebraic in $\tau$. They are actually independent of edge effects discussed in Sec. II.B and we don't need to assume a sufficiently smooth localization function $N(x)$.
 
\begin{figure}
\includegraphics[scale=.5]{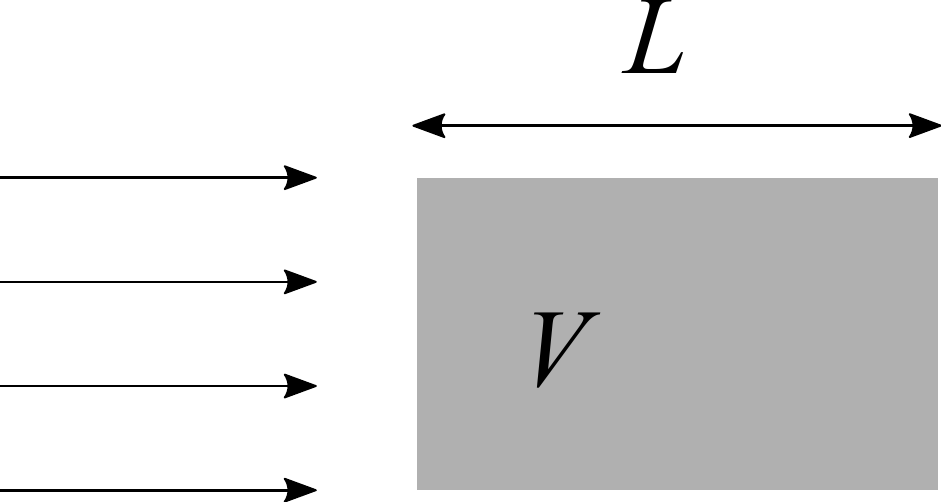}
\caption{A simple measurement of a single particle. The particle goes into the detector's volume $V$ of the length $L$ and is measured over the time $\tau$.}
\label{sing}
\end{figure}

\begin{figure}
    \centering
    \includegraphics[scale=.5]{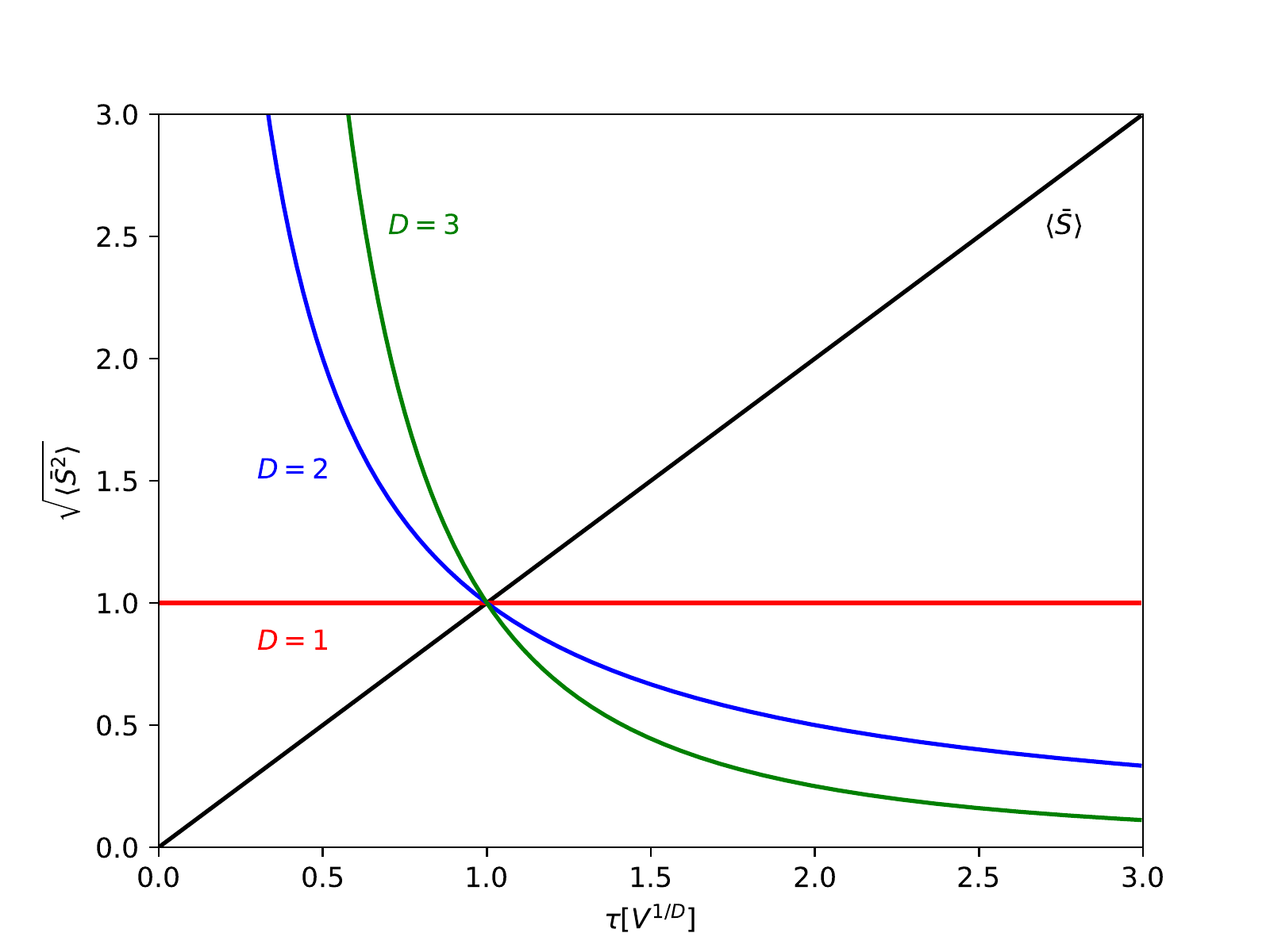}
    \caption{Qualitative comparison between the signal $\langle \bar{S}\rangle$ and vacuum fluctuation $\langle\bar{S}^2\rangle^{1/2}$ for $S=j^3$ and 
a massless particle depending on the dimension $D$ with respect to the measurement time $\tau$.}
    \label{fluct}
\end{figure}

Now, let us consider a single massless particle above the vacuum, with the four-momentum $(E=|k^3|,0,0,k^3)$ . To make it simple, let us take a detector of volume $V$ with periodic boundary condition and the measurement time $\tau$. Then, $\langle \bar{S}\rangle\sim \tau$ for $S=j^0,j^3$ and $\langle \bar{S}\rangle\sim E\tau$ for $S=T^{00}$.
For an open space and the length $L$ along the propagation direction, shown in Fig. \ref{sing},  the measurement time $\tau$ is replaced by $L$ (the speed is $1$). 
There will be no difference if a wavepacket of finite size travels in open space.
Now, the relation between $\langle\bar{S}\rangle$  and $\langle\bar{S}^2\rangle$ will essentially depend on the parameters $\tau$, $L$ and $E$.
For instance, for a very short measurement, the significant contribution of the vacuum fluctuations will dominate the signal, because $\langle\bar S\rangle$ gets smaller while $\langle \bar{S}^2\rangle $ gets larger with decreasing $\tau$, see the qualitative behavior in Fig. \ref{fluct}.

\subsection{Measuring single particles with observables in the  spacelike  spectrum}

\begin{figure}
\includegraphics[scale=.5]{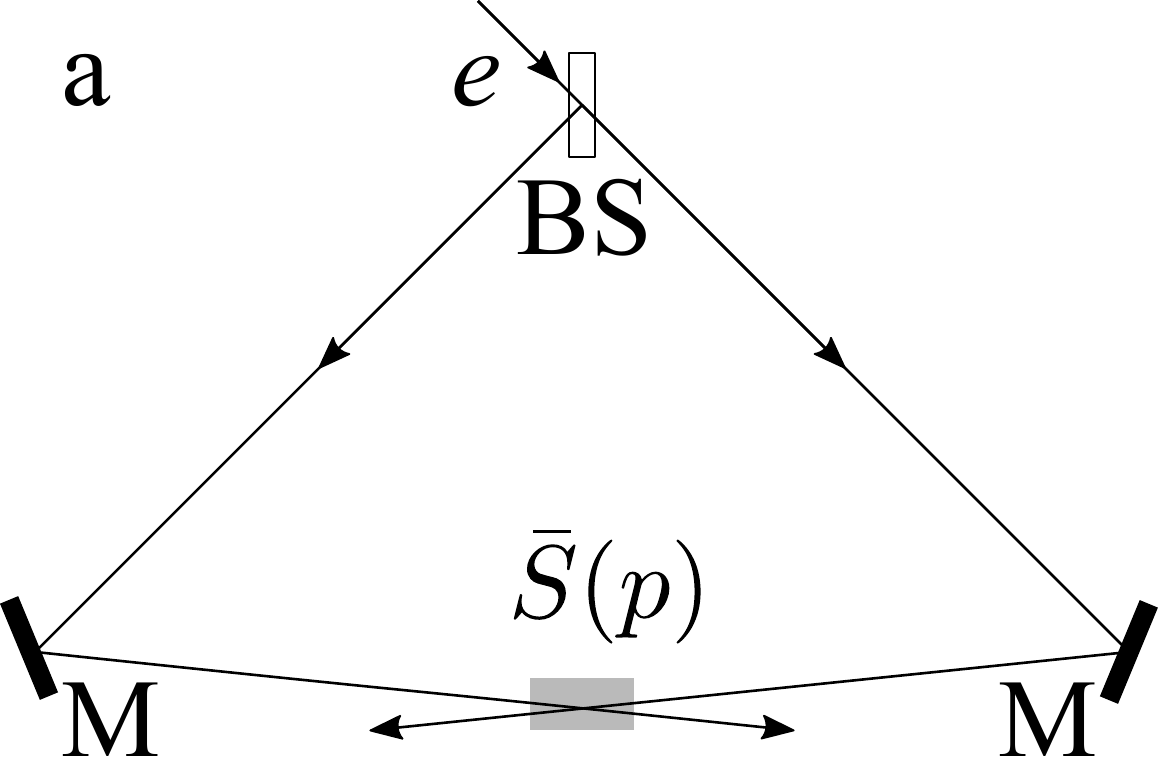}\vspace{1cm}
\includegraphics[scale=.5]{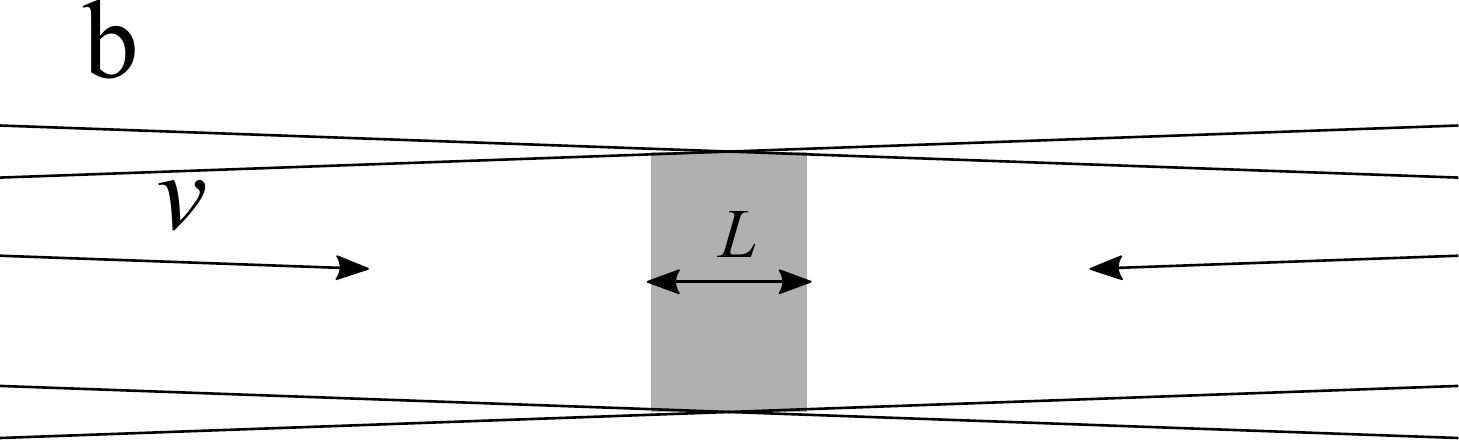}
\caption{Sagnac interferometer configuration to generate
the standing wave. (a) The incoming electron is split into half (beamsplitter BS), with each part bouncing from the mirror (M)
to get counterpropagating and overlapping at the shaded circle, giving nonzero expectation value of $j^\mu(p)$ at spacelike $p$ - momentum difference between
counterpropagating waves. (b) Measurement region  $V$ of counterpropagating waves. The particle travels in the form of the wave packet at the group velocity $v$.
In the overlapping region, we define the rectangle, whose width $L$ along propagation direction, will scale the measurement output.}
\label{sag}
\end{figure}

In the cases where the measurement has a time limitation, the elimination of the vacuum fluctuation terms is desirable.
Hence, we shall employ the lack of spacelike vacuum fluctuations to perform a noiseless measurement of single particles.
The lack of vacuum noise helps to discriminate between registering the particle and no particle.

Let us consider a triangular Sagnac interferometer \cite{sagnac} depicted in Fig. \ref{sag}a.
As we discuss below, in this particular measurement configuration the particle can be detected by observables in spacelike spectrum.
Its actual realization may  depend on the energy of the particles but there is no fundamental obstacle to imagine high-energy counterparts of mirrors and beamsplitters.

We extend the definition (\ref{sss0}) to the observable
\begin{equation}
\bar{S}(p)=\int_{V,\tau} \cos(x\cdot p)S(x)dx.\label{sss}
\end{equation}
restricted to a spatial volume $V$ and a time interval $\tau$ (see Fig. \ref{sag}b).  
Consider now a superposition of two counter-propagating waves of particles of mass $m$, with wave vectors $k=(E,0,0,\pm k^3)$, $E=\sqrt{m^2+(k^3)^2}$ that meet in the triangular loop in Fig. \ref{sag}, and take $p=(0,0,0,2k^3)$ in (\ref{sss}). Then  $\langle\bar{S}(p)\rangle $ for will essentially depend on the
interference of the two waves. To adjust the measured observable to the actual wavepackets, we need to know the measuring region $V$, at least approximately.
For the circular beam cross section, we define it as a cylinder with the axis along the beam, and we take the time window sufficiently long for the particle to pass through it.

The calculation of $\langle \bar{S}(p)\rangle$ depends on the specific type of particle.
We show in Appendix \ref{appg}, assuming the counter-propagating particles and the detector fill all the space $V$ and the measurement lasts the time $\tau$, that
\begin{equation}
    \langle \bar{S}^n(p)\rangle=\langle \bar{S}(p)\rangle^n\,.\label{mom}
\end{equation}
This means the superposition of counter-propagating particles is an eigenstate of $\bar{S}(p)$.
For an open space, we repeat the reasoning in the preceding subsection for the wave group velocity $v=k^3/E$ and the energy $E=\sqrt{m^2+(k^3)^2}$. For a size $L$ (shown in Fig. \ref{sag}b) along the propagation direction the measurement time $\tau$ is replaced by $L/v$. 
It is important that the vacuum noise does not contribute to fluctuations of $\bar{S}(p)$, which helps to avoid dark counts. The remaining contribution from  edge effects of the beam and $V$ can be eliminated by taking sufficiently smooth $N(x)$, as in Sec. II.B.


We can apply the above scenario to the cases of fermions and bosons.
For the Dirac spinor we take the counterpropagating superpositions of left-handed and right-handed states (see Appendix \ref{appg}),
\begin{eqnarray}
&&\sqrt{2}|\psi_a\rangle=|L,k^3\rangle+|R,-k^3\rangle,\nonumber\\
&&\sqrt{2}|\psi_b\rangle=|L,k^3\rangle-|L,-k^3\rangle
\end{eqnarray}
and measure either $S_a=j^0$ or $S_b=j^1$.
One can alternatively also replace $L$ with $R$, reversing the middle sign in the second case.
Then  $\langle\bar{S}_a(p)\rangle_a=\tau m/2E$ and $\langle\bar{S}_b(p)\rangle_b=\tau k^3/2E$. In both cases the average of  $\bar{S}(p)$ is nonzero whereas the noise if given by Eq.~\ref{mom} which allows to detect single fermions not obscured by vacuum fluctuations. However, $S_a$ and $S_b$ dominate at low and high energies, $|k^3|\ll m$ and $|k^3|\gg m$, respectively.  One has to keep coherence between waves, i.e., $S$ must be adjusted if a possible phase shift occurs. 


Next, for a scalar (bosonic) field $\phi$ of mass $m$ we consider the energy-momentum tensor (Appendix \ref{appe})
\begin{equation}
T^{\mu\nu}=\partial^\mu\phi\partial^\nu\phi-g^{\mu\nu}(g^{\sigma\tau}\partial_\sigma\phi\partial_\tau\phi-m^2\phi^2)/2.\label{t00}
\end{equation}
Here the counterpropagating waves with $p=(0,0,0,2k^3)$  give only nonzero $\langle T^{00}\rangle$ and taking $S=T^{00}$ we get $\langle \bar{S}\rangle=\tau m^2/2E$. At high energies it becomes smaller but still should exceed residual vacuum fluctuations caused by edge effects.

A more interesting case is for electromagnetic field $A^\mu=(A^0,\g{A})$, where the tensor reads
\begin{equation}
T^{\mu\nu}=g^{\mu\nu}F_{\delta\gamma}F^{\delta\gamma}/4-F^{\mu\alpha}g_{\alpha\beta}F^{\nu\beta}
\end{equation}
with $F_{\mu\nu}=\partial_\mu A_\nu-\partial_\nu A_\mu$.
Decomposing into electric and magnetic fields $\g{E}=-\partial_t A^0-\nabla \g{A}$, $\g{B}=\nabla\times\g{A}$
we have $T^{00}=(|\g{E}|^2+|\g{B}|^2)/2$ (energy density), $T^{0i}=S^i=(\g{E}\times\g{B})$ (Poynting vector)
and $T^{ij}=E^i E^j+B^iB^j-\delta_{ij}(|\g{E}|^2+|\g{B}|^2)/2$ (Maxwell stress tensor).

Let us consider again two superposition of counterpropagating vertical photons (field $\g{E}$ in positive direction $1$ at $x=0$) with momenta $\g{k}=(0,0,k^3)$ and $-\g{k}$ with $k^3>0$,
\be
\sqrt{2}|\psi\rangle=|V,+k^3\rangle+|V,-k^3\rangle
\ee
Then $\langle \bar{S} \rangle=E\tau/2$ (here $E=|k^3|$) for $S=T^{11}+T^{00}$ or $S=(T^{11}-T^{22})/2$.

The above examples show that the noiseless observables indeed can be used to register single particles
if one overcomes technical problems such as high energy beamsplitters and maintaining coherence.

\subsection{Homodyne detector}

The above configuration suffices to prepare the state and well-defined noiseless observable.
The actual detection must differ a lot from the standard, absorptive chambers \cite{pdg}. Instead, the interference region
must be probed by yet another beam of particles.
For spacelike $p=(0,0,0,2k^3)$, we can detect $j^\mu(p)$ by measuring electromagnetic field 
$A^\mu(p)=-j^\mu(p)/p\cdot p$. We can send a beam of electrons (or other charged particles) of momentum $\g{k}=(0,0,k^3)$ towards the region of 
nonzero $j^\mu(p)$. Then a (small) part of the beam will be reflected with momentum $-\g{p}$ and (partly) changed spin.
For $|k^3|\ll m$ (mass of an electron) we can use a nonrelativistic approximation for single electrons of charge $e$,
\begin{equation}
H=-|\nabla -ie\g{A}|^2/2m+e\g{\sigma}\cdot\g{B}/2m
\end{equation}

\begin{figure}
\includegraphics[scale=.5]{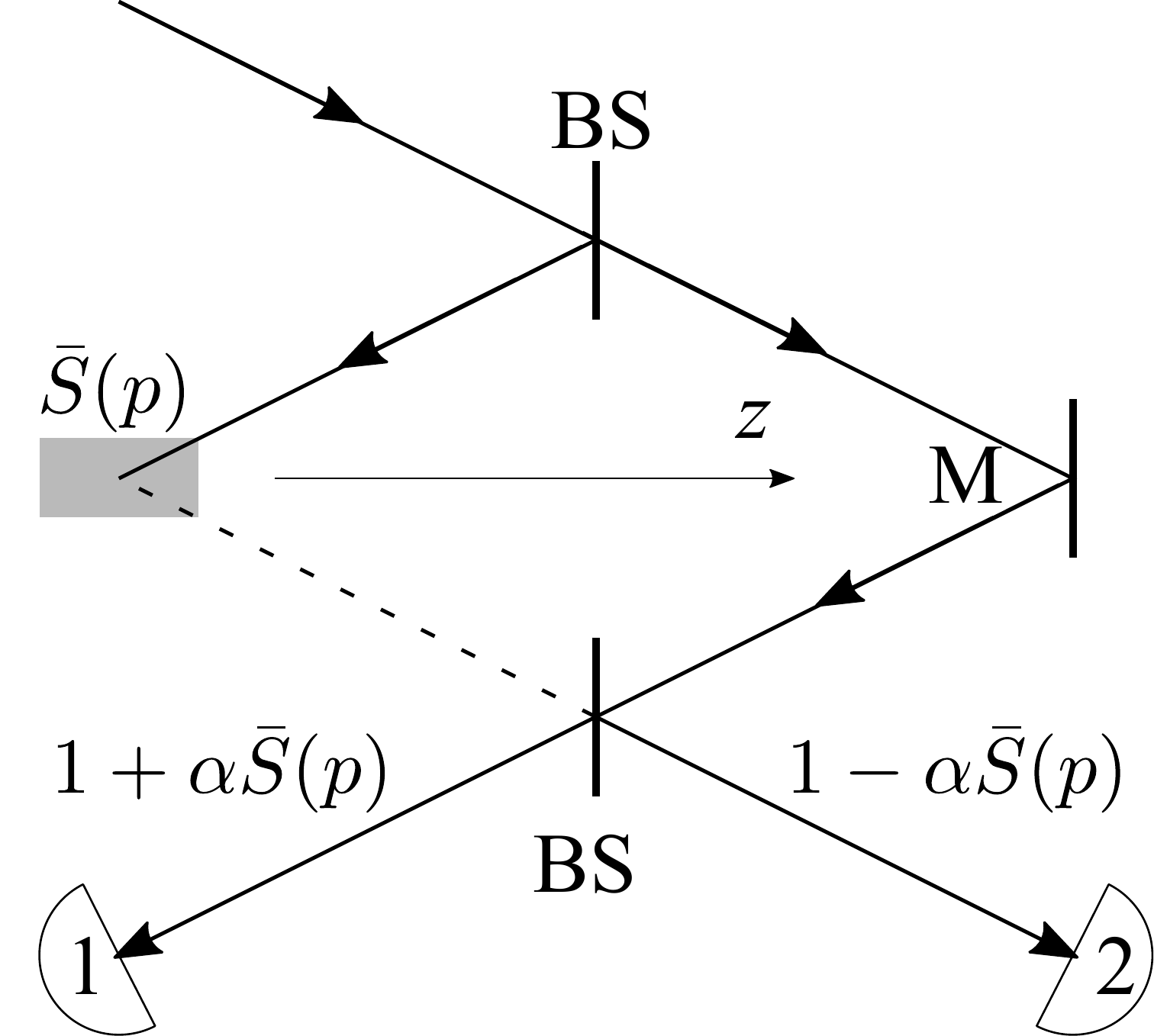}
\caption{Balanced homodyne detection scheme. The electron beam starts from the left upper corner.
It splits on the upper beamsplitter (BS). The upper beam gets partially reflected by interaction with the noiseless observable  $\bar{S}(p)$  (shaded rectangle - the same as in Fig. \ref{sag})
and recombines with the lower beam, reflected by the mirror (M), at the lower beamsplitter. The small
difference between intensities $|1\pm \alpha \bar{S}|^2$ of the beam will be proportional to the value of the noiseless observable.
The scheme can be adjusted by phase shifts and spin rotations if necessary.}
\label{hom}
\end{figure}

We propose a balanced homodyne detection scheme, Fig. \ref{hom} \cite{homod}. Namely, the initial beam with the momentum $-\g{k}$ splits into 
a superposition of left and right arm with momentum $-\g{k}$ and $+\g{k}$. The lower part is reflected by the mirror to get the momentum $-\g{k}$
The left arm beam interacts with $\bar{S}(p)$ for $p=(0,0,0,2k^3)$
(more precisely $\g{B}=\nabla\times A$, which translates into $i\g{p}\times\g{A}$ for $\g{A}$ restricted to $xy$ part)
and some part (proportional to $\bar{S}$ for $S(p)=j^{1,2}(p)$) of the beam gets reflected with momentum $-\g{k}$. Finally, another
beamsplitter (half-splitting) combines the beams so that the left arm gets the amplitude $\sim 1+\alpha \bar{S}(p)$ while the right one  $\sim 1-\alpha \bar{S}(p)$ with some interaction coefficient $\alpha$ to be calibrated 
by the actual overlap of the beam with the region of nonzero $S(x)$ and beam attenuation. The detectors collect particles
and the average difference, proportional to $|1+\alpha \bar{S}(p)|^2-|1-\alpha \bar{S}(p)|^2\simeq 4\alpha \bar{S}(p)$. Additional simple elements (e.g. magnets or movable mirrors) can tune the phase of the particles and calibrate the signal.
On the other hand, the beams should be shielded from stray electromagnetic fields that can cause decoherence.

The scheme can be in principle adopted also to measure $T^{\mu\nu}$ but the naturally coupled gravitational field
is extremely weak. A more realistic approach  would require effective nonlinear electrodynamics, i.e., Euler-Heisenberg term of Lagrangian density \cite{euler}.
 Recent experiments on photon-photon scattering \cite{light} show that this route may be feasible although still demanding, 
with the necessity to extract the right tensor from detector-system interaction given by Euler-Heisenberg Lagrangian, 
possibly with several independent detectors. An alternative is a nonlinear media (crystal) allowing one to probe second-order observables 
(e.g. products of fields) \cite{stein11,stein}.
 
 \subsection{Condensed matter applications}
 
Because of the limited feasibility of the above high-energy proposals, the scheme may be applied to various 1D and 2D condensed matter systems, with spectral relations analogous to relativity. In particular, systems such as electronic leads, junctions, wires, also in the quantum Hall regime can be treated by 1D massless Dirac Hamiltonian. One only needs to replace the speed of light with the Fermi velocity, whose value is material-dependent and much smaller than the speed of light.
The states in the 1D-case reduce to left- and rightgoing electrons with 
 \be
 \hat{H}=\int (\hat{\psi}^\dag_L\partial_1\hat{\psi}_L-\hat{\psi}^\dag_R\partial_1\hat{\psi}_R(x))idx^1
 \ee
for position $x^1$ and $\{\hat{\psi}_{L,R}^\dag(x^1),\hat{\psi}_{L,R}(y^1)\}=\delta(x^1-y^1)$ and zero otherwise. Here the Fermi velocity determines the propagation speed just like the speed of light in vacuum dynamics. In momentum space it reads
\be
 \hat{H}=\sum_k k(\hat{\psi}^\dag_{Rk}\hat{\psi}_{Rk}-\hat{\psi}^\dag_{Lk}\hat{\psi}_{Lk})
\ee
with $\{\hat{\psi}_{L,R,k}^\dag,\hat{\psi}_{L,R,k}\}=1$. At zero temperature, all states $L$ with $k<0$ and $R$ with $k>0$. The case $D=2$ can be realized in graphene \cite{graphene} or topological insulators \cite{kane}. Then
\be
 \hat{H}=\int  \begin{pmatrix}
\hat{\psi}^\dag_L&\hat{\psi}^\dag_R
\end{pmatrix}\begin{pmatrix}
\partial_1&\partial_2\\
\partial_2&-\partial_1\end{pmatrix}
\begin{pmatrix}\hat{\psi}_L\\
\hat{\psi}_R\end{pmatrix}idx^1dx^2
\ee
with $\hat{\psi}$ depending on $x^1,x^2$. The 3D case is also realized in condensed matter resulting in full Dirac dynamics \cite{dirac3da,dirac3db,dirac3dc}. We refrain here from a further analysis of the feasibility of the setup, because it depends on many material-connected factors. Nevertheless, the outlined scheme could be useful well beyond high-energy physics.

\section{Conclusion}

The vacuum noise can limit the accuracy and reliable time scales of detection of single particles at high energies. In order to go beyond this limitation, we proposed to use observables in  the space-like spectrum because they are essentially noiseless in the zero-temperature limit. This lack of noise is a universal property of relativistically invariant systems, following either from the fluctuation-dissipation theorem or the weak positivity (positive definite second order correlations). 

A practical  detection of the particle is then realized by splitting a wavepacket into a standing wave of counter-propagating modes. The detector-system interaction can be used in the homodyne scheme when a beam in one arm of the interferometer interacts weakly with the space-like observable so that the presence of the particle shows up as a difference in final beams' intensities. This general concept can be realized in many ways. Electrons and other charged particles can be probed by another charged beam while photons need the weak photon-photon interaction. From a practical point of view, the proposal relies on the feasibility of beamsplitters or mirrors for high-energy particles or fields and the actual form of the particle-detector interaction. The range of potential implementations will be limited by certain experimental obstacles. Nevertheless, we believe that the fundamental benefit of such a measurement, which is the lack of noise and compliance with relativistic symmetries, will motivate experiments to translate our scheme into a real setup. 

Finally, the concept we presented is not restricted to relativistic high-energy physics. On the contrary, we expect that it will be easier implemented in various condensed matter analogs of relativistic physics.

\section*{Acknowledgement}
A.B. thanks P. Chankowski for many fruitful and inspiring discussions about the topic.

\appendix
\section{Closed time path formalism in relativity}
\label{appa}

We use the following, common conventions and notation in relativistic quantum field theory.
We use dimensionless units i.e., speed of light, Planck's and Boltzmann's constants are $c=\hbar=k_B=1$.
The fourposition $x=(x^0=t,\boldsymbol{x})=(x^0,x^1,..,x^D)$ in 1 time dimension $t$, and $D$ dimensional spatial position $\boldsymbol x$ (in the standard space $D=3$). We identify scalar as without index, e.g. mass $m$, fourvectors by a single index, e.g. $A^\mu$
and tensors with double index $B^{\mu\nu}$. If not ambiguous, indices will be omitted or replaced by the
 summation convention $X^{\mu}_{\mu}=X=\sum_{\mu=0}^{D}X^\mu_\mu$, extending to multiindex expressions. We assume flat metric 
\begin{equation}
g_{\alpha\beta}=g^{\alpha\beta}=\left\{\begin{array}{ll}
+1&\mbox{ for }\alpha=\beta=0\\
-1&\mbox{ for }\alpha=\beta=1..D\\
0&\mbox{ otherwise}\end{array}\right.
\end{equation}
and $X_\alpha=g_{\alpha\beta}X^\beta$, $X\cdot Y=X_\alpha Y^\alpha$, $X^\mu Y_{\mu\nu}Z^\nu=X\cdot Y\cdot Z$.
$\partial_\alpha=\partial/\partial x^\alpha$,
Fourier (energy-momentum) representation of fields (functions of fourposition)
$X(p)=\int e^{ix\cdot p}X(x)dx$ with the integral over the whole spacetime $dx=dx^0\cdots dx^D$. Quantum states and operators can change under Lorentz
transformations, algebraic representations of Lorentz (Poincare in general) group of linear transformations of spacetime, preserving the metric.
 
In general interacting theories, it is more convenient to use an equivalent path integral approach, as it is manifestly compliant with the relativistic symmetries. The correlations (\ref{ccc}) are expressed in terms of path correlations
\begin{eqnarray}
&&\langle A(x)B(y)C(z)\cdots\rangle=\label{ccp}\\
&&Z^{-1}\int D\phi A(x)B(y)C(z)\cdots
\exp\int i\mathcal L(w)dw \nonumber
\end{eqnarray}
where $L$ is the Lagrangian density in terms of local fields $\phi(x)$, while $A$, $B$, $C$ are also functions of local fields. The integral is normalized to
\begin{equation}
Z=\int D\phi \exp\int i\mathcal L(x)dx \label{ccn}
\end{equation}
The integration over $x^{1..D}$ extends over infinite volume, while time $x^0$ flows 
 over Keldysh-Schwinger-Kadanoff-Baym-Matsubara closed time path (CTP) \cite{keldysh,schwinger,kaba,matsubara}  $t(s)$ with $s\in [s_i,s_f]\subset\mathbb R$
with $dt/ds\neq 0$ and $\mathrm{Im}\;dt/ds\leq 0$. $t(s_i)-t(s_f)=i\beta$, $\beta=1/T$ (inverse temperature). In the case $T\to 0$ we have $t(s_\mp)\to \pm i\infty$. The order of $A$, $B$, $C$ must be preserved on the time path. The order will be denoted by $x>y$ if $s_x>s_y$ and then $x>y>z$ in (\ref{ccp}).
In particular, the time can go forward (on real axis), then backwards, and again if necessary.
In most cases a single flat (Keldysh-Schwinger) part suffices which splits into $t\to t_{\pm}=t(s_\pm)=t\pm i\epsilon$ ($\epsilon\to 0_+$, a small positive number going to $0$ in the limit) and $x_\pm=(t_\pm,\boldsymbol x)$ with $s_+<s_-$.
In the case of zero temperature $T\to 0_+$ or $\beta\to +\infty$ the time extends to $\pm i\infty$, see Fig. \ref{ctp}.
The quantum closed time path framework consistent with relativity is summarized in Appendix \ref{appa}.
The correlations in interacting theories can be perturbatively expressed in terms of free propagators and vertices, see \cite{weldon,chou,landsman,kapusta} and Appendix \ref{appb}

\begin{figure}
\includegraphics[scale=.5]{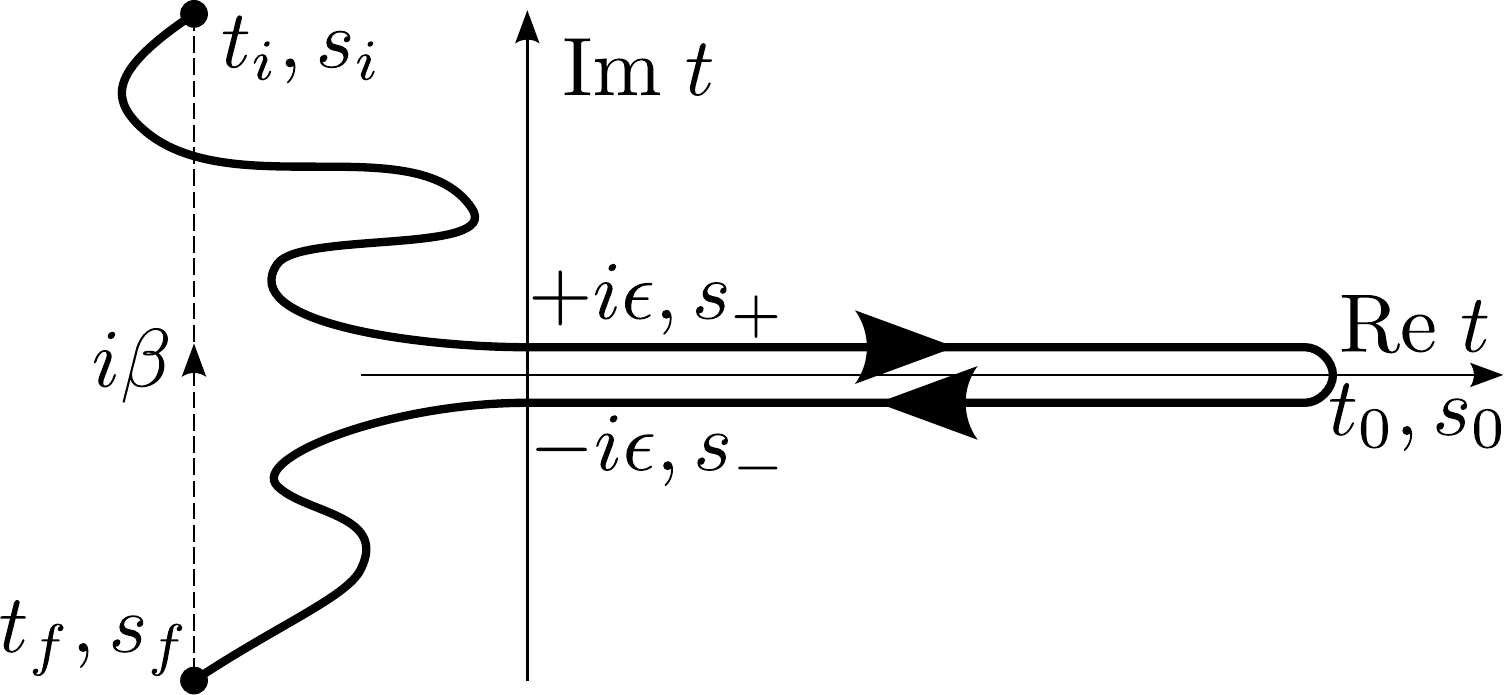}
\caption{The time path in the CTP approach in the case of finite temperature $\beta=1/T$. At zero temperature, the shift $\beta$ stretches to infinity
with $t_i\to +i\infty$, $t_f\to -i\infty$ }\label{ctp}
\end{figure}

Although the Lagrangian density is manifestly relativistically invariant, time has still a special role in CTP.
Fortunately, one can show directly, but perturbatively, that correlations (\ref{ccp}) are invariant at zero temperature,
by direct application of generators of relativistic transformations \cite{ab13}.

The physical interpretation of correlations $\langle A(x)B(y)C(z)\cdots\rangle$ with real $x,y,z,...$ is problematic because it depends on the order on the CTP, i.e.,
 $x^0,y^0,z^0\to x^0,y^0,z^0\pm i\epsilon$. For the two-point correlation $\langle A(x)B(y)\rangle$ one can take $x^0\to x^0+i\epsilon$, $y^0\to y^0-i\epsilon$ so that $x>y$ on the contour. It can be generalized to three or more points using a generalized CTP, see Fig. \ref{ctpx}, so that $x>y>z>...$ on the contour. We shall denote correlations as $\langle \overleftarrow{ \cdots}\rangle$ in this case.
Another important ordering is symmetric time-ordered, common in weak measurements \cite{bfb,ab16}, i.e., one takes the average over $A(x)$
with $x^0\to x^0\pm i\epsilon$, i.e., upper and lower parts of the contour, for all observables. This case will be denoted as $\langle \cdot \rangle_c$.
The last option is full symmetrization of the ordering, i.e., $\langle A_1\cdots A_n\rangle_s=\sum_P\langle\overleftarrow{ P(A_1\cdots A_n)}\rangle/n!$,
summing over all permutations.
It has no connection to any physical measurement model but is free from some artifacts of weak measurements, such as time symmetry violation or energy nonconservation \cite{bfb,energy}. In situations where the order is irrelevant or the differences between orderings are negligible, we will simply use
$\langle\cdot \rangle$ notation. Note that average $\langle A\rangle$ never depends on the order while second order correlations satisfy
$
\langle AB\rangle_c=\langle AB\rangle_s
$

The states, dynamics, and operators in relativistic quantum field theory can be formally defined just like in the nonrelativistic case, in appropriate Hilbert space. However, they are initially defined in a particular reference frame,
so the invariance must be proven. The invariance is much more manifest when translating the standard 
Hamiltonian-based matrix-operator products into correlations with respect to path integrals involving Lagrangian density. 

In the operator approach, we define time-independent field Hermitian operators $\hat{A}(\boldsymbol x)$ which can have additional 
vector or tensor structure and Hamiltonian $\hat{H}$. Next, the time-dependent (Heisenberg) field operators read
$\hat{A}(x)=e^{i\hat{H}x^0}\hat{A}e^{-i\hat{H}x^0}$. The vacuum state is the ground eigenstates of $\hat{H}$ with the lowest possible eigenvalues -- energy. Thermal states are given by $\hat{\rho}=Z^{-1}e^{-\beta \hat{H}}$ for $\beta=1/T$
($T$ - temperature), with $Z=\mathrm{Tr}e^{-\beta\hat{H}}$ so that $\mathrm{Tr}\hat{\rho}$ is normalized to $1$.
The accordance of quantum field theory with relativity requires that its dynamics (defined by Lagrangian density) and the zero-temperature vacuum state are invariant under the Poincare group.
This is formally a Wightman postulate  and it is not trivial to show that a particular model satisfies it.
Fourposition can be transformed according to the Poincare group, which combines translations $x^\mu\to x^\mu+a^\mu$ for
constant fourvector $a$ and Lorentz group rotations $x^\mu\to \Lambda^{\mu}_{\:\nu}x^\nu$ for a constant matrix $\Lambda$
such that $g_{\alpha\beta}=g_{\mu\nu}\Lambda^\mu_{\:\alpha}\Lambda^\nu_{\:\beta}$, so that 
$(x-y)\cdot(x-y)$ is invariant under this group. In fact, here we only need its continuous subgroup, i.e., 
$\det\Lambda=1$ and $\Lambda^0_{\:0}>0$ (excluding time $x^0\to -x^0$ and space $\boldsymbol x\to-\boldsymbol x$ reversal). 
The representations of the Poincare group apply to all quantities appearing
in quantum field theory, including fields (scalar, vector, spinor), states (vacuum, thermal, perturbed), dynamics
(Lagrangian, energy and momentum).
The zero-temperature state, i.e., $\beta\to+\infty$, is relativistic invariance, either by postulates \cite{wightman,jost,coleman} or by direct proof \cite{ab13}

\section{Closed time path propagators for interacting fields}
\label{appb}

The most convenient free Hamiltonian is the quadratic form of bosonic and fermionic operators
\be
\hat{H}_0=\sum_{kl}(b_{kl}\hat{x}_k\hat{x}_l+f_{kl}\hat{\phi}_k\hat{\phi}_l)
\ee
where $\hat{x}$ and $\hat{\phi}$ are Hermitian operators
with bosonic and fermionic commutation relations, respectively,
$[\hat{x}_k,\hat{x}_l]=i g_{kl}\hat{1}$, $\{\phi_k,\phi_l\}=h_{kl}\hat{1}$, with real $g$ and $h$.

One can always diagonalize $\hat{H}_0$ so that
\be
\hat{H}_0=H_{vac}+\sum_{k}E_k(\hat{A}^\dag_k\hat{A}_k+\hat{\psi}^\dag_k\hat{\psi}_k)
\ee
where $H_{vac}$ is the vacuum energy (can be ignored), $\hat{A}$ and $\hat{\psi}$ are linear combinations of $\hat{x}$ and $\hat{\phi}$, respectively,
with the property 
\be
[\hat{A}_k,\hat{A}_l]=\{\hat{\psi}_k,\hat{\psi}_l\}=0,\:[\hat{A}_k,\hat{A}^\dag_l]=\{\hat{\psi}_k,\hat{\psi}^\dag_l\}=\delta_{kl}.
\ee
and $[\hat{A}_k,\hat{\psi}_l]=[\hat{A}^\dag_k,\hat{\psi}_l]=0$.
It is especially simple and instructive to find Green's functions for the above Hamiltonian, extended to the whole CTP.
In the case of bosonic operators, $\langle \h{A}(t)\rangle_0=\langle \h{A}^\dag(t)\rangle_0=0$ and
\begin{eqnarray} 
&&\langle \h{A}_k(t)\h{A}_l(t')\rangle_0=\langle \h{A}^\dag_k(t)\h{A}^\dag_l(t')\rangle_0=0,\label{green}\\
&&\langle  \h{A}_k(t)\h{A}^\dag_l(t')\rangle_0=\delta_{kl} \f{e^{i(t'-t)E_k}}{1-e^{-\beta E_k}}\nonumber\\
&&\langle \h{A}^\dag_k(t)\h{A}_l(t')\rangle_0=\delta_{kl} \f{e^{i(t-t')E_k}}{e^{\beta E_k}-1}.\nonumber\\
&&\langle \h{\psi}_k(t)\h{\psi}_l(t')\rangle_0=\langle \h{\psi}^\dag(t)_k\h{\psi}^\dag_l(t')\rangle_0=0,\nonumber\\
&&\langle \h{\psi}_k(t)\h{\psi}^\dag_l(t')\rangle_0=\delta_{kl}\f{e^{\rmi(t'-t)E_k}}{1+\e^{-\beta E_k}},\nonumber\\
&&\langle \h{\psi}^\dag_k(t)\h{\psi}_l(t')\rangle_0=-\delta_{kl}\f{e^{i(t-t')E_k}}{\rme^{\beta E_k}+1}.\nonumber
\end{eqnarray}
for $t>t'$ on CTP.

The many-point Green's functions are obtained from Wick theorem \cite{wick}. For products of odd number of operators,
the Green's function vanishes while for the even number $2n$,
\ba
&&\left\langle \mathcal T\prod_{k=1}^{2n}  \h{x}_k(t_k)\right\rangle_0=\nonumber\\
&&\frac{1}{n!2^n}\sum_\sigma
\prod_{k=1}^n\langle \mathcal T \h{x}_{\sigma(2k-1)}(t_{\sigma(2k-1)})\h{x}_{\sigma(2k)}(t_{\sigma(2k)})\rangle_0,\nonumber\\
&&\left\langle \mathcal T\prod_{k=1}^{2n} \h{\phi}_k(t_k)\right\rangle_0=\\
&&\frac{\mathrm{sgn}\,\sigma}{n!2^n}\sum_\sigma
\prod_{k=1}^n\langle \mathcal T\h{\phi}_{\sigma(2k-1)}(t_{\sigma(2k-1)})\h{\phi}_{\sigma(2k)}(t_{\sigma(2k)})\rangle_0\nonumber
\ea
with $\mathcal T$ denoting time ordering on CTP, i.e.,
\be
\mathcal T\prod_{k=1}^{2n}  \h{x}_k(t_k)=\h{x}_{\rho(1)}(t_{\rho(1)})\cdots\h{x}_{\rho(2n)}(t_{\rho(2n)})
\ee
where $\rho$ is such a permutation that $t_{\rho(k)}>t_{\rho(j)}$ if $k<j$ and

\be
\mathcal T\prod_{k=1}^{2n}  \h{\phi}_k(t_k)=\mathrm{sgn}\rho\h{\phi}_{\rho(1)}(t_{\rho(1)})\cdots\h{\phi}_{\rho(2n)}(t_{\rho(2n)})
\ee

\section{Proof of exponential suppression of spacelike correlations}
\label{appc}

We shall take for granted that free theories for a scalar, Dirac, or electromagnetic field, have well-defined energy eigenstates with the relativistic constraint $E\geq |\g{P}|$ where $\g{P}$ is the total momentum of the state. Moreover, the eigenstates of
energy are also eigenstates of total momentum, $|\g{P}\rangle$ and any local observable $\hat{X}(\g{x})$ can be decomposed into
\be
\h{X}(\g{x})=\int d\g{p}d\g{q}\hat{X}_{pq}e^{i(\g{p}-\g{q})\cdot\g{x}}|\g{p}\rangle\langle\g{q}|
\ee
where $\hat{X}_{\g{p}\g{q}}$ is $\hat{X}$ restricted to the elements between momentum eigenstates $\g{q}$ and $\g{p}$ (there can be many).
from spacetime invariance we can calculate
\be
G_{XY}(p)=(2\pi)^{D+1}\int dx e^{ip\cdot x}\langle X(x)Y(0)\rangle
\ee
We shall write down $X$ and $Y$ in the eigenbasis of free theory, namely
\be
G_{XY}= (2\pi)^{2D+1}\int dt e^{ip^0 t}\int d\g{q}\langle \hat{X}_{\g{q},\g{q}-\g{p}}(t)\hat{Y}_{\g{q}-\g{p},\g{q}}\rangle
\ee
in the Heisenberg picture.
Finally in the thermal state, with $\hat{\rho}\propto \exp(-\beta\h{H})$ we have
\ba
&&G_{XY}(p)=(2\pi)^{2D+2} Z^{-1}\int d\g{q} e^{-\beta E(\g{q})}\times\nonumber\\
&&\delta(p^0-E(\g{q}-\g{p})+E(\g{q}))\mathrm{tr}\hat{X}_{\g{q},\g{q}-\g{p}}\hat{Y}_{\g{q}-\g{p},\g{q}}
\ea
with the normalization factor $Z$, independent of $p$.

From inequalities
\be
E(\g{q}-\g{p})\geq |\g{q}-\g{p}|\geq |\g{p}|-|\g{q}|
\ee
and $E(\g{q})\geq |\g{q}|$, and restriction $p^0=E(\g{q}-\g{p})-E(\g{q})$, we have
\ba
&&2E(\g{q})=E(\g{q})+E(\g{q}-\g{p})-p^0\nonumber\\
&&\geq |\g{q}|+|\g{p}|-|\g{q}|-p^0\geq|\g{p}|-|p^0|
\ea
In this way the damping factor is $\leq e^{(|p^0|-|\g{p}|)\beta/2}$.

For interacting theories, we have to insert all combinations of $\h{H}_I$ perturbatively.
Without loss of generality we insert it $k$ times after $\h{Y}$ and before $\h{X}$ and $n$ times
after $\h{X}$ and before $\h{Y}$, including the vertical Matsubara part, contributing to $n+k$th perturbation order
(ignoring temporarily corrections to global normalization $Z$).
We decompose
\be
\h{H}_I=\sum_{\g{p}}\hat{H}_{I\g{p}}
\ee
with $\h{H}_{I\g{p}}$ restricted to momentum eigenstates $|\g{p}\rangle$.
We assume the dynamics to be translation invariant so $\h{H}_I$ cannot mix different momenta.
We shall additionally label momentum eigenstates with their (noninteracting) energy, i.e., $|E\g{p}\rangle$.
We are left with the following integral contributing to $G_{XY}$
\ba
&&\int dt \mathcal T\int_t^{-i\beta} d^nt'
e^{iE'_n(t'_n+i\beta)}\langle E'_0\g{q}|\h{X}_{\g{q},\g{q}-\g{p}}|E''_{k}\g{q}-\g{p}\rangle
\nonumber\\
&&\left(\prod_{r=0}^{n-1}e^{iE'_r(t'_{r}-t'_{r+1})}
\langle E'_{r+1}\g{q}|\h{H}_{I\g{q}}| E'_{r}\g{q}\rangle\right)
\times\\
&&\mathcal T\int_0^t d^kt'' e^{iE''_k(t''_k-t)}e^{ip^0 t}\langle E''_0\g{q}-\g{p}|\h{Y}_{\g{q}-\g{p},\g{q}}|E'_{n}\g{q}\rangle\nonumber\\
&&\left(\prod_{j=0}^{k-1} 
e^{iE''_{j+1}(t''_{j}-t''_{j+1})}
\langle E''_{j+1}\g{q}-\g{p}|\h{H}_{I\g{q}-\g{p}}|E''_j\g{q}-\g{p}\rangle\right)\nonumber
\ea
Here $d^nt'=dt'_1\cdots dt'_n$, $d^kt''_k=dt''_1\cdots dt''_k$, $\mathcal T$ denotes time order, i.e., $t''_{j+1}>t''_j$ and $t'_{r+1}>t'_r$ with the order along CTP, and
$t''_{0}=0$, $t''_{k}<t$, $t'_{0}=t$, $t'_{n}<-i\beta$.

The critical  factor is the integral over times, which we reordered into
\ba
&&\int dt \mathcal T\int_t^{-i\beta} d^nt'
e^{iE'_n(t'_n+i\beta)}
\prod_{r=0}^{n-1}
e^{iE'_r(t'_{r}-t'_{r+1})}\times\nonumber\\
&&
\mathcal T\int_0^t d^kt'' e^{iE''_k(t''_k-t)}e^{ip^0 t}
\prod_{j=0}^{k-1} 
e^{iE''_{j}(t''_{j}-t''_{j+1})}
\ea
equal
\ba
&&\int dt e^{ip^0 t}e^{-\beta E'_n}e^{i(E'_0-E''_k)t}\mathcal T\int_t^{-i\beta} d^nt'\times\\
&&\prod_{r=1}^{n}
e^{i(E'_r-E'_{r-1})t'_r}
\mathcal T\int_0^t d^kt''
\prod_{j=1}^{k} 
e^{i(E''_{j}-E''_{j-1})t''_{j}}.\nonumber
\ea
The integrals over $t'$ and $t''$ can be done recursively. For instance
\be
\int_0^{t''_2} dt''_1e^{i(E''_{1}-E''_{0})t''_{1}}=\f{e^{i(E''_1-E''_0)t''_2}-1}{i(E''_1-E''_0)}
\ee
Of course it can happen that $E''_1=E''_0$ but then the integral has a well defined limit.
We can prove by induction that
\be
\int_0^{t''_{s+1}} d^s t''\prod_{j=1}^{s} 
e^{i(E''_{j}-E''_{j-1})t''_{j}}=\sum_{j=0}^{s} A^s_j e^{i(E''_s-E''_j)t''_{s+1}}
\ee
with coefficients $A^s_j$ independent of $t''_{s+1}$ (but depending on $E''$).
By induction hypothesis the integral reads
\ba
&&\sum_{j=0}^{s-1}A^{s-1}_j \int_0^{t''_{s+1}}dt''_s e^{i(E''_{s}-E''_{s-1})t''_{s}}e^{i(E''_{s-1}-E''_j)t''_{s}}
\nonumber\\
&&=\sum_{j=0}^{s-1}A^{s-1}_j\f{e^{i(E''_s-E''_j)t''_s}-1}{i(E''_s-E''_j)}
\ea
so $A^s_j=A^{s-1}_j/i(E''_s-E''_j)$ for $j<s$ and $A^s_s=\sum_{j=0}^{s-1}i/(E''_s-E''_j)$
Therefore
\be
\mathcal T\int_0^t d^kt''
\prod_{j=1}^{k} 
e^{i(E''_{j}-E''_{j-1})t''_{j}}=\sum_{j=0}^{k} A_j e^{i(E''_k-E''_j)t}
\ee
and, shifting time,
\ba
&&e^{-\beta E'_n}e^{i(E'_0-E''_k)t}\mathcal T\int_t^{-i\beta} d^nt'
\prod_{r=1}^{n}
e^{i(E'_r-E'_{r-1})t'_r}=\nonumber\\
&&\sum_{r=0}^{n} B_r e^{-iE''_kt}e^{E'_n(it-\beta)}e^{i(E'_n-E'_r)(-i\beta-t)}
\ea
so finally we get the factor
\be
\sum_{j=0}^k\sum_{r=0}^nA_jB_r\int dt e^{ip^0 t} e^{i(E'_r-E''_j)t-\beta E'_r}
\ee
Integrating over $t$ we get
\be
\sum_{j=0}^k\sum_{r=0}^n2\pi A_jB_r\delta(p^0+E'_r-E''_j) e^{-\beta E'_r}
\ee
We have then
\be
2E'_r=E'_r+E''_j-p^0\geq |\g{q}|+|\g{q}-\g{p}|-p^0\geq |\g{p}|-p^0
\ee
which gives the same estimate as in the noninteracting theory.

The above proof ignored global corrections to normalization $Z$. 
We can incorporate these corrections by  perturbative expansion in two-point propagators (not just vertices). However,  the problem is that not only normal-ordered propagators
$\langle \hat{A}^\dag\h{A}\rangle$ or $\langle \h{\psi}^\dag\h{\psi}\rangle$, with positive energy appear in the expansion, but also
antinormal propagators, i.e.,  $\langle \hat{A}\h{A}^\dag\rangle$ or $\langle \h{\psi}\h{\psi}^\dag\rangle$, with \emph{negative} energy, i.e., $-E_k$. Fortunately, they are damped by $e^{-\beta E_k}$, see (\ref{green}). We shall see that
a careful collecting of these damping factors will restore eventually the same global damping as in the previous proof.

\begin{figure}
\includegraphics[scale=.5]{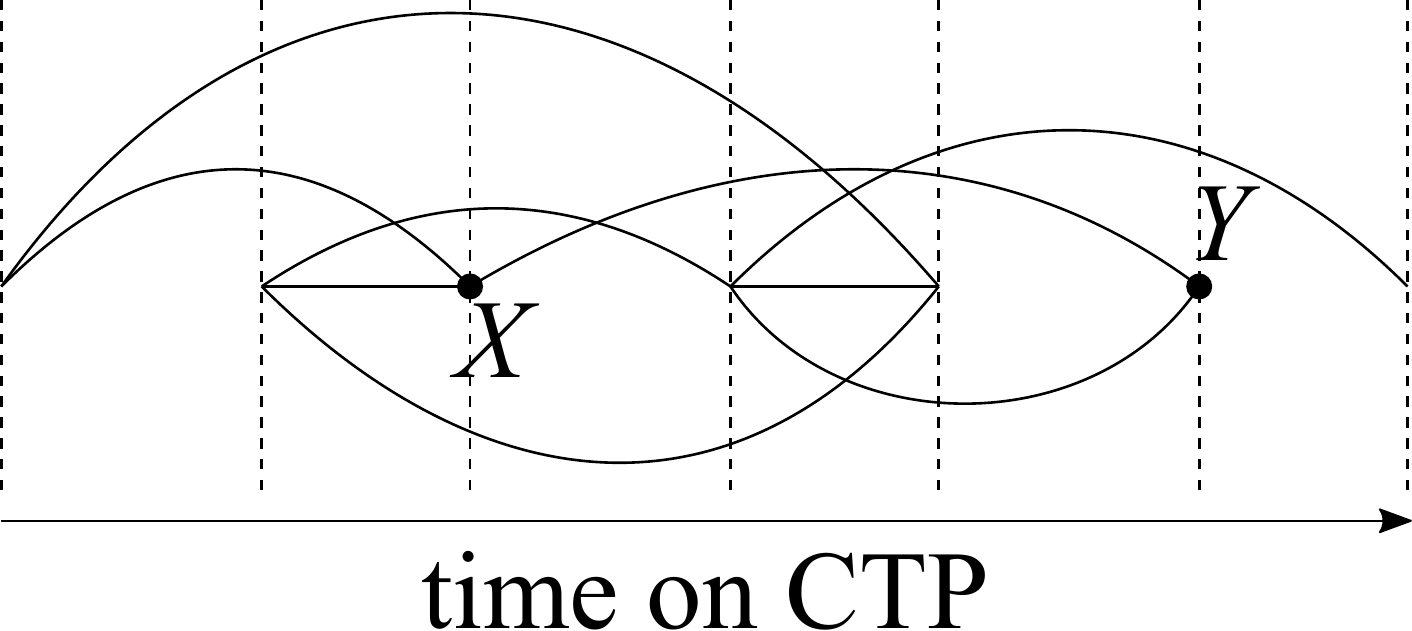}
\caption{The propagators in the proof of decay of correlations are cut at all times corresponding to some vertices $H_I$,
even for those that do not end at this $H_I$
}

\label{tcut}
\end{figure}

As in the previous proof, we can time-order $\h{H}_I$ and integrate recursively over times. It does not matter that only some
propagators have their endpoints at specific $\h{H}_I$, we just integrate \emph{all} propagators
whose  time interval (specified by endpoints) contains the given time interval, see Fig. \ref{tcut}. It applies also to the vertical
shift by $-i\beta$. The energies $E'_r$ and $E''_j$ are now sums of individual energies of the propagators, i.e.,
\be
E'_r=E'_{r+}-E'_{r-},\:E''_j=E''_{j+}-E''_{j-}
\ee
with
\be
E'_{r\pm}=\sum_\alpha \epsilon'_{r\alpha\pm},\:
E''_{j\pm}=\sum_\gamma \epsilon''_{j\gamma\pm}
\ee
where the $\alpha+$ and $\gamma+$ indicate positive energies, while $\alpha-$ and $\gamma-$ indicate negative energies
(each $\epsilon$ and $E_\pm$ is positive by definition, the actual sign is given explicitly in front of the energy)
The negative energies bring the damping factor $e^{-\beta E'_{r-}}$, $e^{-\beta E''_{j-}}$
 (some energies $\epsilon_-$ can cover several intervals but it is here irrelevant, we pick them only once).
The final expression contains the sum
\be
\sum_{j=0}^k\sum_{r=0}^n2\pi A_jB_r\delta(p^0+E'_r-E''_j)e^{-\beta E'_r-\beta E'_{r-}-\beta E''_{j-}}
\ee
Now $E'_r=E''_j-p^0$ so $E'_{r+}+E''_{j-}=E'_{r-}+E''_{j+}-p^0$ and
\ba
&&2(E'_r+E'_{r-}+E''_{j-})=2(E'_{r+}+E''_{j-})\nonumber\\
&&=E'_{r+}+E'_{r-}+E''_{j+}+E''_{j-}-p^0
\ea
On the other hand, momentum conservation (each propagator has a well defined momentum $\g{p}$ that sums up to the total momentum) and the fact that $\epsilon(\g{p})\geq |\g{p}|$ for individual momentum $\g{p}$, gives
\be
E'_{r+}+E'_{r-}\geq |\g{q}|,\: E''_{j+}+E''_{j-}\geq |\g{q}-\g{p}|
\ee
completing the proof.

\section{Restriction of vacuum correlations}
\label{appd}

The Lorentz invariance implies that $G^{\mu\nu}(p)=p^\mu p^\nu \eta(p\cdot p)-g^{\mu\nu}\xi(p\cdot p)$
with some functions $\xi$ and $\eta$ \cite{ab15,ab16}. Let us check the positivity of $G$ for $p=(0,...,0,q)$.
Then $G^{11}=\xi=-G^{00}$ so $\xi=0$.

A bit more complicated reasoning applies to a generic symmetric tensor quantity $B^{\mu\nu}(x)=B^{\nu\mu}(x)$
and $B^{\mu\nu}(p)=\int dx e^{ip\cdot x}B^{\mu\nu}(x)$. Then
\begin{equation}
\langle B^{\mu\nu}(x)B^{\sigma\rho}(y)\rangle=G^{\mu\nu\sigma\rho}(x-y)
\end{equation}
from translation invariance and
\begin{equation}
\langle B^{\mu\nu}(p)B^{\sigma\rho}(q)\rangle=(2\pi)^{D+1}\delta(p+q)G^{\mu\nu\sigma\rho}(p)
\end{equation}
Lorentz invariance and symmetry implies that
\begin{eqnarray}
&&G^{\mu\nu\sigma\rho}(p)=p^\mu p^\nu p^\sigma p^\rho a-p^\mu p^\nu g^{\sigma\rho}b-p^\sigma p^\rho g^{\mu\nu} b^\ast
+\nonumber\\
&&(p^\mu p^\sigma g^{\nu\rho}+p^\mu p^\rho g^{\nu\sigma}+p^\nu p^\sigma g^{\mu\rho}+p^\nu p^\rho g^{\mu\sigma})f\nonumber\\
&&
+(g^{\mu\sigma}g^{\nu\rho}+g^{\nu\sigma}g^{\mu\rho})v+g^{\mu\nu}g^{\sigma\rho}w
\end{eqnarray}
where $a,f,v,w$ are real functions of $p\cdot p$. The exception is $b$ that can be complex and depend also on the sign of $p^0$ but only for $p\cdot p>0$, with $b^\ast(p)=b(-p)$, while $b$ is real for $p\cdot p<0$.

We shall analyze positivity of $G$ only for $D\geq 3$ and $p\cdot p<0$. The $(D+1)\times (D+1)$ matrix has some symmetries.
Let us take $p=(0,0,0,q)$. Then $G^{1212}=v=-G^{0101}$ so $v=0$ and $G^{0303}=q^2f=-G^{1313}$ so $f=0$ while $a$, $b$ (now only real) and $w$ combine to a positive definite
form
\begin{eqnarray}
&&G^{\mu\nu\sigma\rho}=(g^{\mu\nu}-(b/w)p^\mu p^\nu)(g^{\sigma\rho}-(b/w)p^\sigma p^\rho)w\nonumber\\
&&+p^\mu p^\nu p^\sigma p^\rho(a-b^2/w)
\end{eqnarray}
with $w\geq 0$ and $aw\geq b^2$. We obtain (\ref{gbb}) by replacing $b/w\to b$ and $a-b^2/w\to a$. 

For an antisymmetric tensor $B^{\mu\nu}=-B^{\nu\mu}$ (e.g. field $F^{\mu\nu}=\partial^\mu A^\nu-\partial^\nu A^\mu$) the generic invariant correlation reads
\begin{eqnarray}
&&G^{\mu\nu\sigma\rho}(p)=\epsilon^{\mu\nu\rho\sigma}a+(g^{\mu\sigma}g^{\nu\rho}-g^{\nu\sigma}g^{\mu\rho})v+\\
&&(p^\mu p^\sigma g^{\nu\rho}-p^\mu p^\rho g^{\nu\sigma}-p^\nu p^\sigma g^{\mu\rho}+p^\nu p^\rho g^{\mu\sigma})f\nonumber
\end{eqnarray}
where $\epsilon^{\mu\nu\rho\sigma}$ is a constant completely antisymmetric tensor. For spacelike $p$, taking again $p=(0,0,0,q)$ we get
$G^{1212}=v=-G^{0101}$ so $v=0$, $G^{0303}=q^2f=-G^{1313}$ so $f=0$, $G^{0123}=a$ so $a=0$.

\section{Scalar field and its energy-momentum}
\label{appe}

Real scalar field $\hat{\phi}(\boldsymbol x)$ with conjugate field $\hat{\pi}(\boldsymbol x)$, with commutation relation
\be
[\hat{\phi}(\boldsymbol x),\hat{\pi}(\boldsymbol y)]=i\delta(\boldsymbol x-\boldsymbol y)
\ee
Relativistic field Hamiltonian
\be
\hat{H}=\int d\boldsymbol x(\hat{\pi}^2(\boldsymbol x)+|\nabla \hat{\phi}(x)|^2+m^2\hat{\phi}^2(\boldsymbol x))/2
\ee
Here the $\nabla$ terms is in fact a sum of partial derivatives
\be
|\nabla \hat{\phi}(x)|^2=\sum_{j=1}^3(\partial_j\hat{\phi}(\boldsymbol x))^2
\ee
Heisenberg picture (for $x=(t,\g{x})$)
\be
\hat{\phi}(x)=e^{i\hat{H}t}\hat{\phi}(\boldsymbol x)e^{-i\hat{H}t}
\ee

Translation into path integrals
\be
\langle\Phi'|\exp(-i\hat{H}t)|\Phi\rangle=\int_{\phi(x^0=0,...)=\Phi}^{\phi(x^0=t,...)=\Phi'}\mathcal D\phi \exp\int i\mathcal L(x)dx,
\ee
where
\be
2\mathcal L(x)=\partial\phi(x)\cdot\partial\phi(x)-m^2\phi^2(x)
\ee

Derivative rule: 
\be
\partial_0=(dt/ds)^{-1}\partial/\partial s
\ee
and differential $dx=dx^0dx^1\cdots dx^D$ with $dx^0=(dt/ds)ds$ and delta
\be
\delta(x-y)=\delta(x^0-y^0)\delta(x^1-y^1)\delta(x^2-y^2)\delta(x^3-y^3)
\ee
with $\delta(x^0-y^0)=\delta(s_x-s_y)/(dt/ds)|_{s=s_x=s_y}$.

With such definition one can calculate all relevant quantum field theory functions, i.e.,
\ba
&&\langle \phi(x)\phi(y)\cdots\rangle=\langle \mathcal T\hat{\phi}(x)\hat{\phi}(y)\cdots\rangle=\nonumber\\
&&Z^{-1}\int \mathcal D\phi \phi(x)\phi(y)\exp\int i\mathcal L(z)dz
\ea
with
\be
Z=\int \mathcal D\phi \exp\int i\mathcal L(z)dz
\ee
where $\mathcal T$ denotes ordering by $s$, i.e.,
\be
\mathcal T\hat{\phi}(y)\hat{\phi}(z)=\left\{\begin{array}{ll}
\hat{\phi}(x)\hat{\phi}(y)&\mbox{ if }s_x>s_y\\
\hat{\phi}(y)\hat{\phi}(x)&\mbox{ if }s_y>s_z\end{array}\right.
\ee
Simple correlations read
\be
\langle \phi(x)\phi(y)\rangle=\frac{\int \mathcal D\phi \phi(x)\phi(y)\exp\int i\mathcal L(z)dz}
{\int \mathcal D\phi \exp\int i\mathcal L(z)dz}
\ee
with special cases defined on the flat Keldysh part (Im $x\to 0$).
We denote $\phi_\pm(x)=\phi(x_\pm)$, $2\phi_c(x)=\phi_+(x)+\phi_-(x)$ and
 $\phi_q(y)=\phi_+(y)-\phi_-(y)$, so that $\langle\phi_q(x)\phi_q(y)\rangle=0$
 and

\ba
&&\langle\phi_+(p)\phi_-(q)\rangle
=(2\pi)^5 \delta(q\cdot q-m^2)\theta(q^0)\delta(q+k),\nonumber\\
&&
\langle\phi_c(p)\phi_c(q)\rangle
=16\pi^5\int \delta(q\cdot q-m^2)\delta(q+p),\\
&&
\langle \phi_c(p)\phi_q(q)\rangle=(2\pi)^4\delta(q+p)i/(q_+\cdot q_+-m^2),\nonumber
\ea
 where $q^0_+=q^0-i\epsilon$ ($\epsilon\to 0_+$), in the zero temperature limit.

Energy-momentum stress tensor by Noether theorem
\be
T^{\mu\nu}=\partial^\mu\phi\partial^\nu\phi-g^{\mu\nu}(g^{\sigma\tau}\partial_\sigma\phi\partial_\tau\phi-m^2\phi^2)/2
\ee

where we denoted $T^{\mu\nu}_q(x)=T^{\mu\nu}(x_+)-T^{\mu\nu}(x_-)$

\ba
&&T^{\mu\nu}_c(x)=\partial^\mu\phi_c\partial^\nu\phi_c-g^{\mu\nu}(\partial\phi_c\cdot\partial\phi_c-m^2\phi_c^2)/2
\nonumber\\
&&+\partial^\mu\phi_q\partial^\nu\phi_q/4-g^{\mu\nu}(\partial\phi_q\cdot\partial\phi_q-m^2\phi_q^2)/8
\ea

\section{Higher order correlations}
\label{appf}

\begin{figure}
\includegraphics[scale=.5]{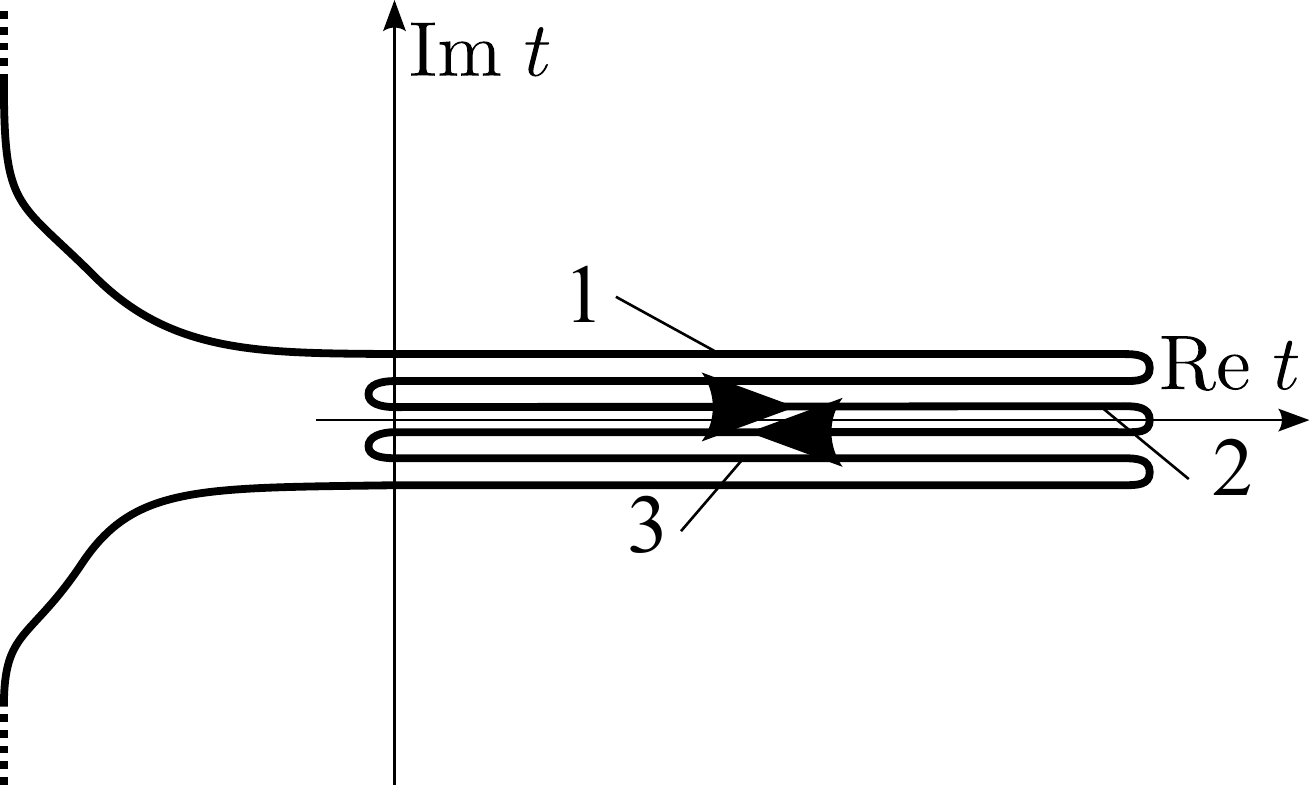}
\caption{CTP modified to incorporate 3 time parts, 1,2,3, in this order on the contour. For better visibility the
parts are separated by $i\epsilon$ with $\epsilon\to 0_+$}
\label{ctpx}
\end{figure}

The most natural definition, consistent with natural notion of weak measurement \cite{chou,bfb}, is symmetrizing on CTP, i.e
\begin{equation}
T^{\mu\nu}\to T^{\mu\nu}_c(x)=(T^{\mu\nu}(x_+)+T^{\mu\nu}(x_-))/2
\end{equation}
with $x_\pm=(x^0\pm i\epsilon,\g{x})$ on the upper/lower flat part of the  contour. No problems occur for averages and second order correlations, but starting from the third order, correlations become problematic.
Let us consider $D=3$ and
\begin{equation}
\langle T^{\mu\nu}(k)\phi(-p)\phi(-q)\rangle=(2\pi)^4\delta(p+q-k)G^{\mu\nu}(p,q)\label{ttq}
\end{equation}
for a real scalar field $\phi$ and $k=(0,0,0,k^3)$, see Figure \ref{tqq}.
For the standard order the correlation reads (see Appendix \ref{appe}),
\begin{equation}
\langle T^{\mu\nu}_c(k)\phi_c(-p)\phi_c(-q)\rangle
\end{equation}

\begin{figure}
\includegraphics[scale=.5]{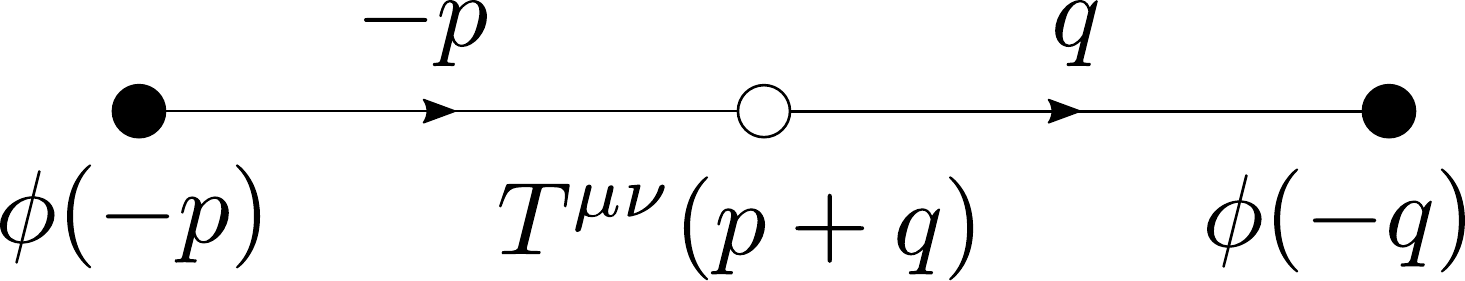}
\caption{The graph contributing to (\ref{ttq}). The lines are $\langle \phi\phi\rangle$ propagators with arrows
indicating the direction of positive momentum (labeled).}
\label{tqq}
\end{figure}

Let us take e.g. $p=(v,0,0,w/2)$, $q=(-v,0,0,w/2)$ so that also $k=(0,0,0,w)$ is definitely spacelike.
Then the only terms contributing to the correlation originate from $\phi_q$ in $T^{\mu\nu}$ giving the expression proportional to
\begin{equation}
\frac{p^\mu q^\nu-g^{\mu\nu}(p\cdot q+m^2)/2}{(p\cdot p-m^2)(q\cdot q-m^2)}
\end{equation}
which is obviously nonzero for $\mu=\nu=3$ ($\sim w^2/8+m^2/2$ for $v=0$). It it also nonzero for the noiseless combination
$2T^{00}+T^{11}+T^{22}$ ($\sim  -2v^2$).

An attempt to resolve this problem, namely symmetrized ordering, i.e.,
splitting CTP into three parts (see Figure \ref{ctpx}) and defining
\begin{equation}
\langle A(x)B(y)C(z)\rangle=\sum_{\sigma(123)}\langle A(x_1)B(x_2)C(x_3)\rangle
\end{equation}
with summation over permutations of $123$, restores conservation, i.e., $\partial_\mu T^{\mu\nu}=0$
but still gives nonzero
\begin{equation}
\langle (2T^{00}(k)+T^{11}(k)+T^{22}(k))\phi(-p)\phi(-q)\rangle
\end{equation}
with $2T^{00}+T^{11}+T^{22}$ being noiseless in the sense of (\ref{noiseo}) and Sec. \ref{sec2a}.
The problematic term is $\phi_1 T_2\phi_3$ with $b$ on the middle part.
The term can be written as
\begin{eqnarray}
&&\int dp'dq'(2p^{\prime 00}q^{\prime 00}+p^{\prime 11}q^{\prime 11}+p^{\prime 22}q^{\prime 22})
\times\\
&&\delta(k-p'-q')\langle \phi_+(-p)\phi_{-}(p')\rangle\langle \phi_+(q')\phi_-(-q)\rangle\nonumber
\end{eqnarray}
Taking $p=(E,0,0,w/2)$, $q=(-E,0,0,w/2)$ for $E=\sqrt{m^2+w^2/4}$ the propagators are nonzero (on-shell delta)
but the prefactor equals $-2E^2$, again nonvanishing. Note that it is necessary to have timelike quantities $\phi(-p)$
with $p\cdot p>0$ in the correlation. Symmetrized correlations of only spacelike momenta will vanish as shown in \cite{bound2}
(alternatively one can adopt our proof in Appendix \ref{appc}).

\section{Calculation  of moments of spacelike measurement for counterpropagating superpositions}
\label{appg}

Introducing Pauli matrices
\ba
&&\sigma^0=\begin{pmatrix}
1&0\\
0&1\end{pmatrix},\:\sigma^1=\begin{pmatrix}0&1\\1&0\end{pmatrix},\nonumber\\
&&
\sigma^2=\begin{pmatrix}0&-i\\i&0\end{pmatrix},\:\sigma^3=\begin{pmatrix}1&0\\0&-1\end{pmatrix}
\ea
we define matrices $\gamma$ ($4\times 4$) in Weyl representation
\be
\gamma^0=\begin{pmatrix}
0&1\\1&0\end{pmatrix},\:\gamma^k=\begin{pmatrix}0&\sigma^k\\-\sigma^k&0\end{pmatrix}
\ee
where $0$ and $1$ denote $2\times 2$ matrices with $0$s and $1$s on the diagonal, respectively.

In the volume $V$ the field reads \cite{peskin}
\be
\hat{\psi}(x)=\sum_{k,X=L,R}\frac{1}{\sqrt{V}}(\hat{a}^X_k u^X_ke^{-ik\cdot x}+\hat{b}^{\dag X}_{k}v^X_{k}e^{ik\cdot x})
\ee
with particle annihilation operators $\hat{a}$ and antiparticle creation $\hat{b}^\dag$ and $E=k^0=\sqrt{m^2+|\g{k}|^2}$.
The anticommutation relations read
\be
\{\hat{a}^X_k,\hat{a}^{\dag X}_k\}=\{\hat{b}^X_k,\hat{b}^{\dag X}_k\}=1
\ee
while all other anticommutators are zero.

Free states of momentum $\g{k}=(0,0,k^3)$ have the general spinor form, with left-handed and right-handed states
\ba
&&u^L_k=(2E)^{-1/2}
\begin{pmatrix}
\sqrt{E-k^3}\theta(-k^3)\\
\sqrt{E+k^3}\theta(k^3)\\
\sqrt{E+k^3}\theta(-k^3)\\
\sqrt{E-k^3}\theta(k^3) \end{pmatrix},\nonumber\\
&&u^R_k=(2E)^{-1/2}
\begin{pmatrix}
\sqrt{E-k^3}\theta(k^3)\\
\sqrt{E+k^3}\theta(-k^3) \\
\sqrt{E+k^3}\theta(k^3)\\
\sqrt{E-k^3}\theta(-k^3) \end{pmatrix}
\ea
 At $k^3=0$ the left-handed and right-handed spinors interchange.
In the limit $m\to 0$  the spinors simplify to
\be
u^L\to 
\begin{pmatrix}
\theta(-k^3) \\
\theta(k^3)\\
0\\
0\end{pmatrix},\;
u^R\to 
\begin{pmatrix}
0\\
0\\
\theta(k^3) \\
\theta(-k^3) \end{pmatrix}
\ee

The current operator $\hat{j}=\hat{\psi}^\dag\gamma^0\gamma\hat{\psi}$ can be simplified to the spinor matrix form
\ba
&&
j^0=\begin{pmatrix}
1&0&0&0\\
0&1&0&0\\
0&0&1&0\\
0&0&0&1\end{pmatrix},\:
j^1=\begin{pmatrix}
0&-1&0&0\\
-1&0&0&0\\
0&0&0&1\\
0&0&1&0\end{pmatrix},
\\
&&
j^2=\begin{pmatrix}
0&i&0&0\\
-i&0&0&0\\
0&0&0&-i\\
0&0&i&0\end{pmatrix},\:
j^3=\begin{pmatrix}
-1&0&0&0\\
0&1&0&0\\
0&0&1&0\\
0&0&0&-1\end{pmatrix}.\nonumber
\ea

Let us evaluate $\langle j^1(p)\rangle$ for $p=(0,0,0,2k^3)$ on the superposition
\be
|\psi\rangle=(|L,k^3\rangle+|R,-k^3\rangle)/\sqrt{2}
\ee
where $|X,k^3\rangle=\hat{a}^{\dag X}_{(E,0,0,k^3)}|0\rangle$ ($|0\rangle$ -- vacuum state).
We find
\be
\langle j^0(p)\rangle_a=\langle \bar{S}_a\rangle_a=\tau m/2E
\ee
for $S_a=j^0$
where $\tau$ is the total measurement time.

For
\be
|\psi_b\rangle=(|L,k^3\rangle-|L,-k^3\rangle)/\sqrt{2}
\ee
we evaluate for $S_b=j^1$
\be
\langle j^1(p)\rangle_b=\langle\bar{S}_b\rangle_b=\tau k^3/2E
\ee
For higher moments, we get simply $\langle \bar{S}^n_{a,b}\rangle=\langle \bar{S}_{a,b}\rangle^n$.

For a scalar field we expand
\be
\hat{\phi}(x)=\sum_{k}\frac{1}{\sqrt{2EV}}(\hat{a}_k e^{-ik\cdot x}+\hat{a}^{\dag }_{k}e^{ik\cdot x})
\ee
with commutators $[\hat{a}_k,\hat{a}^\dag_k]=1$ and all others zero.
Taking the state
\be
|\psi\rangle=(|k^3\rangle+|-k^3\rangle)/\sqrt{2}
\ee
where $|k^3\rangle=\hat{a}^\dag_{(E,0,0,k^3)}|0\rangle$
and $S=T^{00}$ given by (\ref{t00}) we get
\be
\langle\bar{S}^n\rangle=(\tau m^2/4E)^n
\ee

For the electromagnetic field, let use the standard quantization
\be
\hat{\g{A}}(x)=\sum_{k,\lambda}\frac{1}{\sqrt{2EV}}(\g{e}_k^\lambda\hat{a}^\lambda_k e^{-ik\cdot x}+\g{e}^{\ast\lambda}_k\hat{a}^{\dag\lambda }_{k}e^{ik\cdot x})
\ee
with $\lambda=H,V$ for the two polarizations and $E=|\g{k}|$ and commutation $[\hat{a}^\lambda_k,\hat{a}^{\dag\lambda}_k]=1$. In our case let us take 
\be
\g{e}_{(E,0,0,k^3)}=(1,0,0)
\ee
We have $\g{B}=\nabla\times \g{A}$ and $\g{E}=-\partial_0\g{A}$. 
Let us take  the state
\be
|\psi\rangle=(|V,k^3\rangle+|V,-k^3\rangle
\ee
with $|V,k^3\rangle=\hat{a}^{\dag V}_{(E,0,0,k^3)}|0\rangle$.
Then
$
\langle\bar{T}^{00}\rangle=0
$
while
\be
\langle(\bar{T}^{11})^n\rangle=\langle(-\bar{T}^{22})^n\rangle=\tau E
\ee

\end{document}